\documentclass{article}
\usepackage[american]{babel}
\usepackage[latin1]{inputenc}
\usepackage{graphicx}
\usepackage{epsfig}
\usepackage{color}

\begin{document}

\title{Network generation and evolution based on spatial and opinion dynamics components}


\author{Andr\'e C. R. Martins\\
NISC - EACH, Universidade de S\~ao Paulo\\
Av. Arlindo B\'etio, 1000, S\~ao Paulo, 03828-080, Brazil}

\date{}

\maketitle


\begin{abstract}
	In this paper, a model for a spatial network evolution based on a Metropolis simulation is presented. The model uses an energy function that depends both on the distance between the nodes and the stated preferences. The agents influence their network neighbors opinions using the CODA model. That means each agent has a preference between two options based on its probabilistic assessment of which option is the best one. The algorithm generates realistic networks for opinion problems as well as temporal dynamics for those networks. The transition from a random state to an ordered situation, as temperature decreases, is described. Different types of networks appear based on the relative strength of the spatial and opinion components of the energy.
	
\end{abstract}

\section{Introduction}

Most opinion dynamics models are implemented with a fixed neighborhood. That neighborhood defines which agents may interact \cite{castellanoetal07,galam12a,galametal82,galammoscovici91,sznajd00,deffuantetal00,martins08a,martins12b}. Those models are used to study questions such as whether consensus will emerge or if extremism \cite{deffuantetal02a,amblarddeffuant04,galam05,weisbuchetal05,franksetal08a} will prevail in an artificial society. As part of that exploration, the influence of different network structures on the final opinions is a relevant question \cite{Albi2016,Mai2017}. Indeed, it is a well established conclusion different networks might help societies to achieve consensus or dissent \cite{Kurmyshev2011,Acemoglu2011,Das2014,Hu2017}. 

The co-evolution of networks and opinions has been studied for some of the most popular opinion dynamics models \cite{Vazquez2013}. Models for network evolution, however, are usually built with simple probabilistic rules \cite{Zanette2006,Holme2006,Kimura2008,martins08b,Benczik2009,Herrera2011,10.1371/journal.pmed.1002049,Baumgaertner2017}.  More recently, papers with an emphasis on strategies for achieving consensus or synchronization have started appearing \cite{Acosta-Escorcia2015,Dong2017}. But we still lack more sophisticated models that can better describe the evolution of networks. Here, the possible influence of opinions and the location of the agents will be investigated.

Traditional models for generating networks \cite{Erdoes1960,wattsstrogatz98a,Barabasi1999} can account several observed features in general situations. But while those models describe one statistics well, they fail to do the same for other measures of network properties \cite{doi:10.1137/S003614450342480}. Extensions of those models, including rewiring rules, aimed at a better description of relevant variables, have been studied \cite{Krapivsky2002,Vazquez2003,XU20112429,COLMAN201480,10.1371/journal.pone.0187538,RevModPhys.87.925}.
Based on the initial ideas on how to generate Markov graphs \cite{Frank1986}, exponential random graph models~\cite{Wasserman1996,Robins1999,doi:10.1086/427322} have been proposed as a way to generate network models that depend on known or assumed structural features. While those models allow for estimating networks \cite{10.2307/2289546}, it is often the case that those estimates have degeneracy problems \cite{Handcock03assessingdegeneracy}. 

A network type that is particularly interesting for social problems is spatial networks \cite{Waxm1988,Barthelemy2011,Guillier2017,Balister2018}. Interactions often depend on how physically close people are, even if that dependence has become much less important thanks to technological advances. That means agents should have a stronger tendency to link with those who are close than with distant ones. Indeed, some cases echo chambers seem to be influenced by geographical locations. At other times, echo chambers seem to depend only on opinions \cite{Bastos2017}.

Here, the problem of the evolution of opinions on an evolving spatial network is addressed. Rewiring rules to optimize an energy function based on the difference in opinions as well as the distance between the agents are introduced. Indeed, a model based on optimization of objectives related to node and degree edges while keeping the average shortest path length constant was able to generate several different known types of network~\cite{Zheng2014}. 
While studying the problem of designing computers, and, in particular, the travelling salesman problem, Kirkpatrick et al \cite{Kirkpatrick1983} showed we can use simulation techniques, including simple Metropolis algorithms \cite{metropolisetal53a} to solve multivariate or combinatorial optimization problems. Here, we will define an energy function for the network allowing the dynamics to be driven by a simple Monte Carlo update rule.

The opinions will be updated using the Continuous Opinions and Discrete Actions (CODA) model \cite{martins08a,martins12b}. CODA allows us to study problems with binary choices while still measuring the strength of opinion of the agents. It has also suggested a framework for discrete \cite{martins13c} as well as continuous opinion models \cite{martins08c}. Those characteristics allowed the study of problems such as the evolution of extremism \cite{Lu2015}, the emergence of extremism in a society of moderate agents \cite{martinsgalam13a,martins16a}, and the adoption of addictive behaviors \cite{Moore2015,Sun2017}. Effects of different networks of agents for the CODA model and variants have been studied \cite{Deng2013,Chowdhury2016}.  Those effects include the emergence of a trust network as agents update their opinions also on who they trust \cite{martins13b}.

\section{The Model}

\subsection{Opinion updating}

To start, the dynamics of both opinio update and network evolution processes must be defined. For the opinions, the traditional and simplest version of CODA model algorithm \cite{martins08a} will be used. In it, each agent $i$ start with an opinion $\nu_i$ that represents its preference between two choices, $A$ and $B$. When $\nu_i > 0$, agent $i$ prefers $A$, otherwise, it prefers $B$. Agents can only observe the preference of their neighbors, that is, $\sigma_j = sign(\nu_j)$. The update rule when agent $i$ observes the preference $\sigma_j$ of agent $j$ is given by

\begin{equation}\label{eq:opupdate}
\nu_i(t+1) = \nu_i(t) + \sigma_j(t).
\end{equation}

That rule can be obtained from an application of a simplified Bayesian reasoning about the problem. If each agent assumes their neighbor have a fixed probability $\alpha$ they will provide the best answer, we can calculate how they will change their probabilities $p_i$ that $A$ is the best choice by applying Bayes theorem. Equation~\ref{eq:opupdate} can be obtained by estimating the log-odds of $p_i$, $\nu_i = \ln(\frac{p_i}{1-p_i})$, plus a renormalization of $\nu_i$ so that the constant addition term becomes $\pm 1$ \cite{martins08a}. The renormalization works as a counting of how many steps each agent is from changing its preference. The actual size of each step for the log-odds is given by $\ln(\frac{\alpha}{1-\alpha})$.

When there are well defined neighborhoods, that dynamics leads to the appearance of regions supporting each choice . For completely random cases, with no concept of neighborhood, or for fully connected networks, of course, those regions can not be formed. In that case, opinions change randomly until one becomes dominant and drives the system to consensus. 

\begin{figure}
	\centering
	\begin{tabular}{cc}
		\epsfig{file=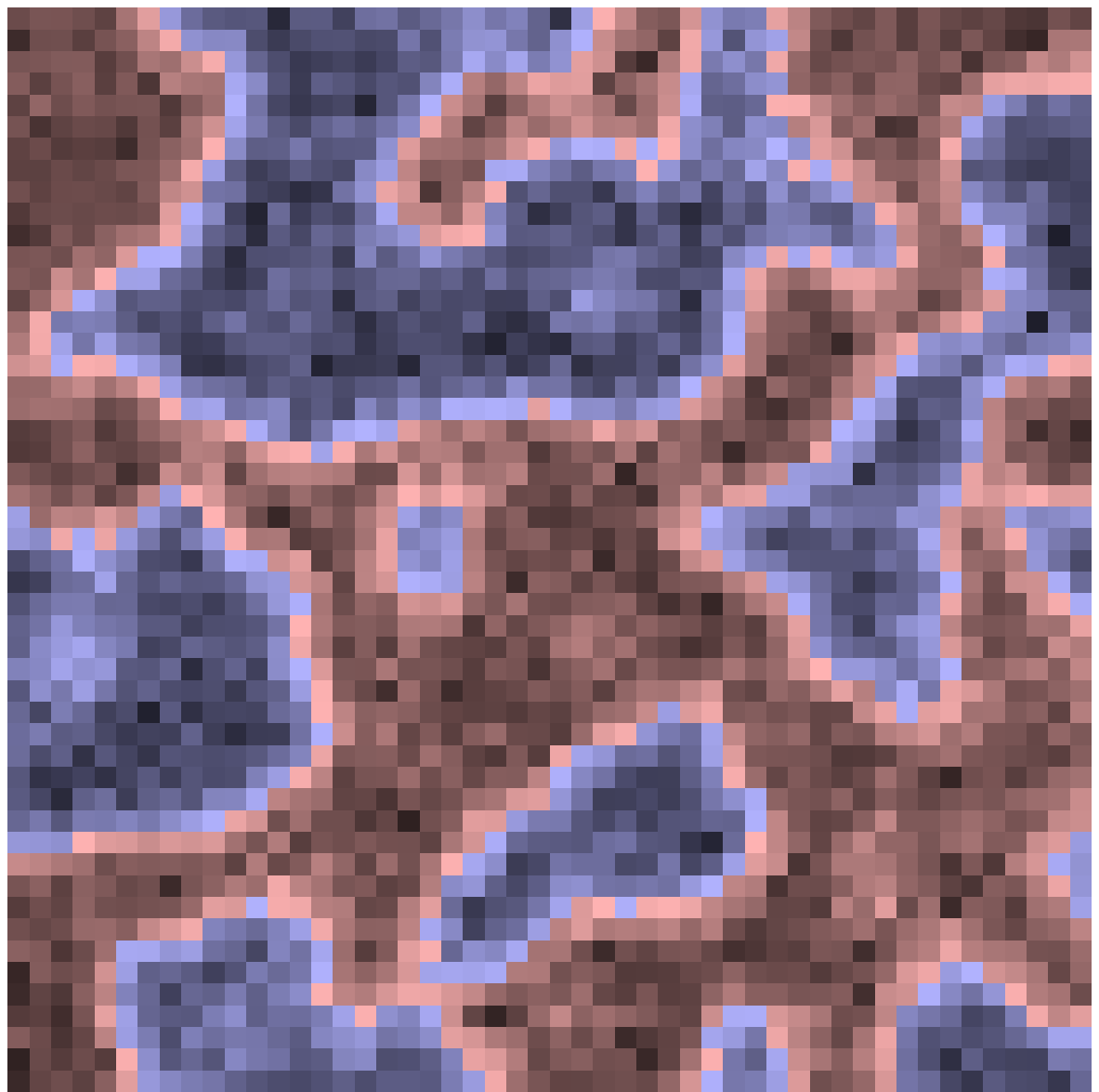,width=0.45\linewidth,clip=} &
		\epsfig{file=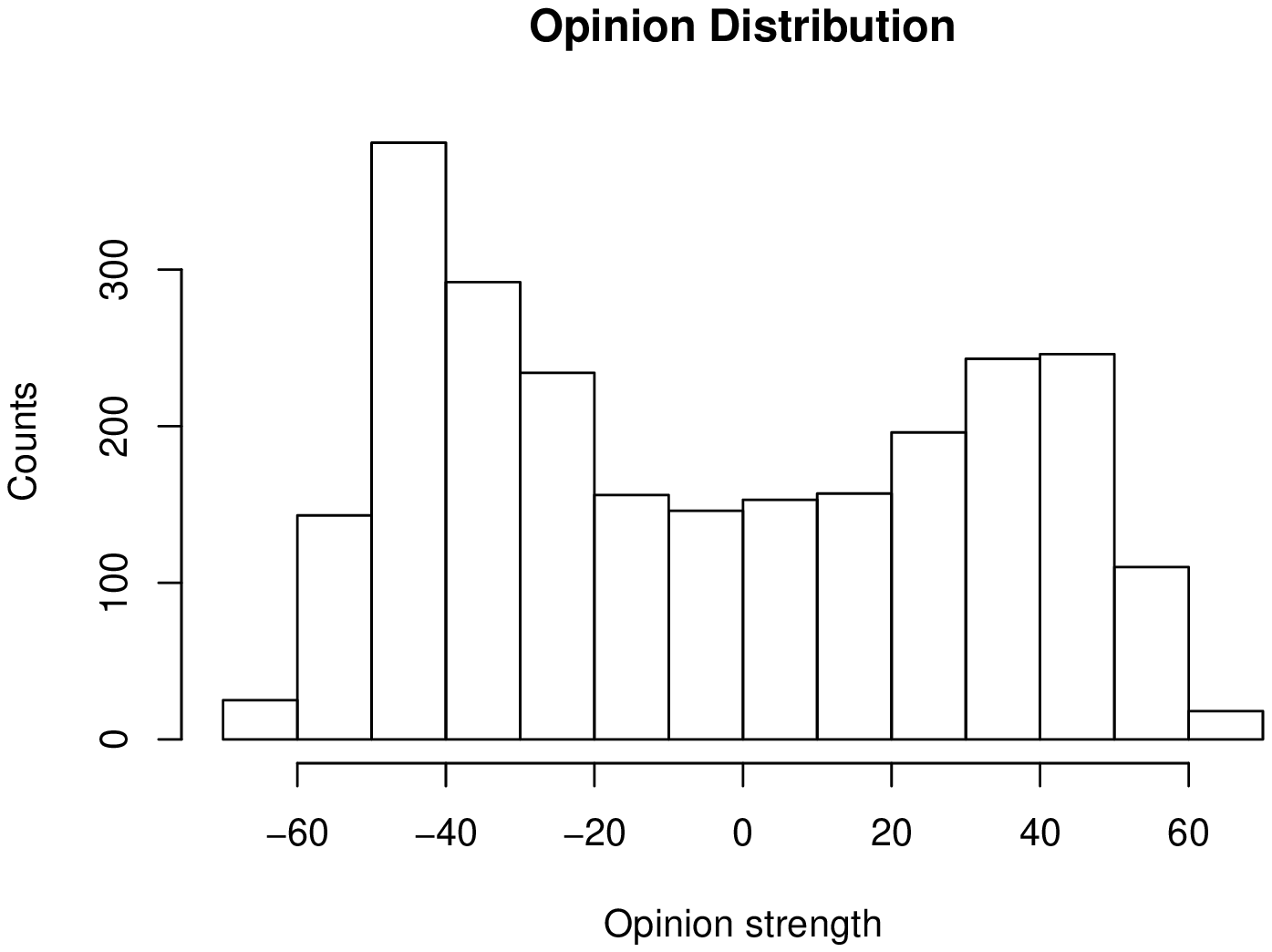,width=0.45\linewidth,clip=}
	\end{tabular}
	\caption{Results of a typical CODA model run over a $50\times50$ square network with empty boundaries. The outcome corresponds to what was observed after an average of 50 interactions per agent. {\it Left} Typical domain configuration. Different preferences are shown as red or blue, with darker shades corresponding to stronger opinions (larger values of $|\nu | $). {\it Right} Histogram showing the strength of opinion $\nu$ measured as number of steps from changing preferences.}\label{fig:coda}
\end{figure}

When neighborhoods exist, that consensus is not observed. Unlike other discrete opinions models, echo-chamber regions soon become well defined. They tend to remain stable as the simulation proceeds. Opinions inside same preference regions reinforce each other to the point that, after a while, few changes in the actual preferences happen. Preference domains freeze, except for small boundary changes. And even those become rarer the longer the simulation lasts. One example of those domains, for a  $50\times50$ square network with empty boundaries, can be seen in the left panel of Figure~\ref{fig:coda}. For that simulation, each agent was influenced by a random neighbor 50 times, in average.

A histogram showing strength of opinion of the agents in the same run can be seen in the right panel. It is easy to notice the two peaks around 50, the average number of interactions per agent. That suggests that an important fraction of the interactions do happen between agreeing, reinforcing neighbors. The central moderate opinions are usually observed at the borders of the domains.

\subsection{Network evolution}

Each agent $i$ will correspond to a specific node (also represented by $i$, for ease of notation) in a spatial network. Each node will exist in a $d$-dimensional space with an Euclidean metric. In the applications and examples that follow, we will usually have $d=2$, with a few $d=1$ examples. That is, each node-agent will have a position associated to it defined either by a single variable $x_i$ or the pair $(x_i , y_i )$. The distance between two nodes, $d_{ij}$ $i$ and $j$ will be the obvious $d_{ij}=|x_i - x_j |$ for $d=1$ or $d_{ij}=\sqrt{(x_i - x_j)^2 +(y_i - y_j)^2}$ for $d=2$.

The initial network edges will be a regular lattice in most cases, with each node connected to all its neighbors up to level $l$. To introduce dynamics for the network structure, an energy function $H$ \cite{Kirkpatrick1983} will be introduced.  $H$ will depend on the distance between connected nodes. Given the set $E$ of all $e$ edges for a network $N$, $H[N]$ will be defined as
\begin{equation}\label{eq:energydist}
H[N] = \beta \sum_E d_{ij}^{\gamma},
\end{equation}
where $\beta$ is, as usual, related to a temperature of the system and $\gamma$ is an exponent introduced to control the effect of very large distances, if needed. In most simulations, we will study the case where $\gamma=1$.

From Equation \ref{eq:energydist}, we can define simple updates rule for the network by using a Metropolis algorithm. Here, we will keep the number of edges $e$ constant. That can be done by randomly picking an existing edge $k$ in $E$ that links two nodes $i_1$ and $i_2$ as well as drawing two unconnected nodes $i_3$  and $i_4$. If we try to replace the link between $i_1$ and $i_2$ for a new link between $i_3$  and $i_4$, the change in energy will be given by
\[
\Delta H = \beta(d_{34}^{\gamma}-d_{12}^{\gamma}).
\]
If the change in energy is negative, the change is automatically accepted. Otherwise, we change the links with a probability given by
\begin{equation}\label{eq:probtransdist}
P=\exp(-\Delta H )= \exp[-\beta(d_{34}^{\gamma}-d_{12}^{\gamma})].
\end{equation}
Of course, other variables could also influence the energy, such as the degree of each node. As the goal of this paper is to study spatial effects, those will not be included. Only the effect of opinions will be added bellow.

\begin{figure}
	\centering
	\begin{tabular}{cc}
		\epsfig{file=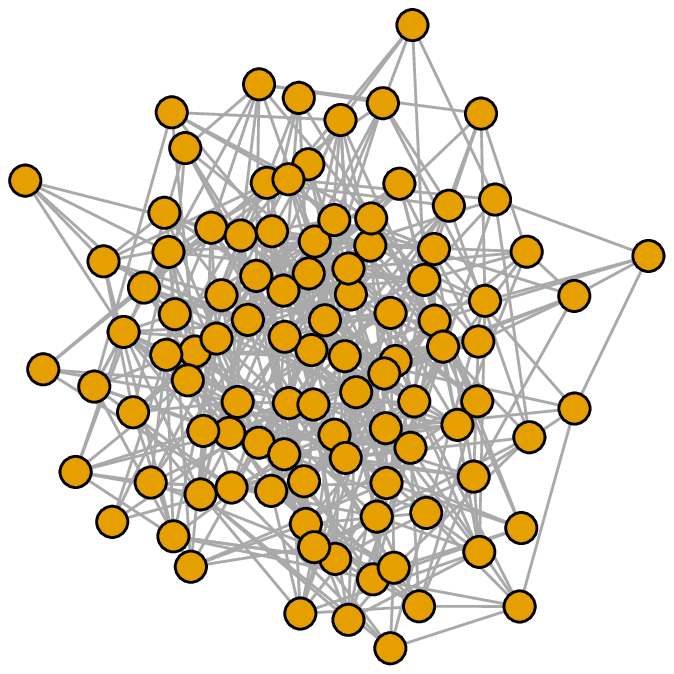,width=0.45\linewidth,clip=} &
		\epsfig{file=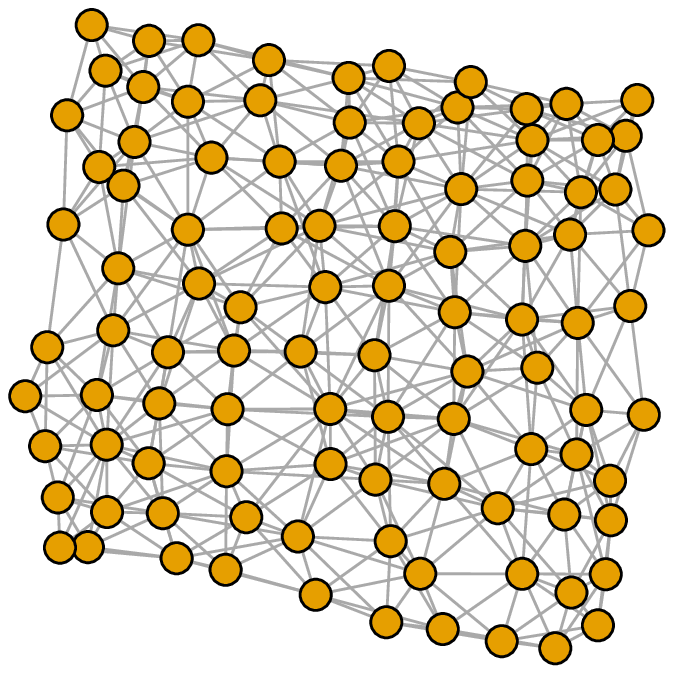,width=0.45\linewidth,clip=}
	\end{tabular}
	\caption{Networks generated after 100,000 interactions with $\gamma=1$, starting with a $d=2$, $10\times 10$ lattice with second neighbors connected and empty boundary conditions. {\it Left} $\beta=0.2$. {\it Right} $\beta=10.0$.}\label{fig:simplenetworks}
\end{figure}

Figure \ref{fig:simplenetworks} shows two examples of networks generated by the algorithm after 100,000 attempted updates. In those examples, the energy depends linearly with the distance, that is, $\gamma=1$. In both cases, the initial network was a regular $d=2$ lattice with second neighbors connected and empty boundary conditions. The networks were drawn using the igraph package \cite{Csardi2006} running in the R software environment \cite{Rsoftware}. The left panel shows the results for a higher temperature ($\beta=0.2$), while the right panel corresponds to a lower one ($\beta=10.0$). It is easy to see that when $\beta$ is large, the systems moves to a state close to its minimum energy. That is, closer nodes tend to be linked while distant ones do not. As the temperature increases (smaller $\beta$), termal agitation wins and we observe a more random network.

To include opinion effects, agents will also consider the preferences of their neighbor on a certain issue. That is, agents will have a tendency to link to other agents who share their same preference. That effect can be introduced in our network update algorithm by adding a term that depends on the choices $\sigma$ of the agents. Since agents prefer agreement (unless one is trying to model contrarians \cite{galam04,martinskuba09a}), we can add to the energy a term proportional to $-\sigma_i \sigma_j$ . That means Equation \ref{eq:energydist} becomes
\begin{equation}\label{eq:energydistop}
H[N] = \beta \sum_E (d_{ij}^{\gamma} - J \sigma_i \sigma_j)
\end{equation}
where $J$ measures the relative importance of distance and opinion effects on the energy. The probability of accepting a transition to a higher energy state  at Equation \ref{eq:probtransdist} becomes
\begin{equation}\label{eq:probtransdistop}
P = \exp[-\beta(d_{34}^{\gamma}-d_{12}^{\gamma}- J \sigma_i \sigma_j)].
\end{equation}

In real world, observing the opinion of others could be more common than rewiring one's network. To investigate if that possibility has an effect on the network statistics, each time the algorithm tries a rewire will correspond to $k$ opinion updates.

\section{Simulation results}

\begin{figure}
	\centering
	\begin{tabular}{cc}
		\epsfig{file=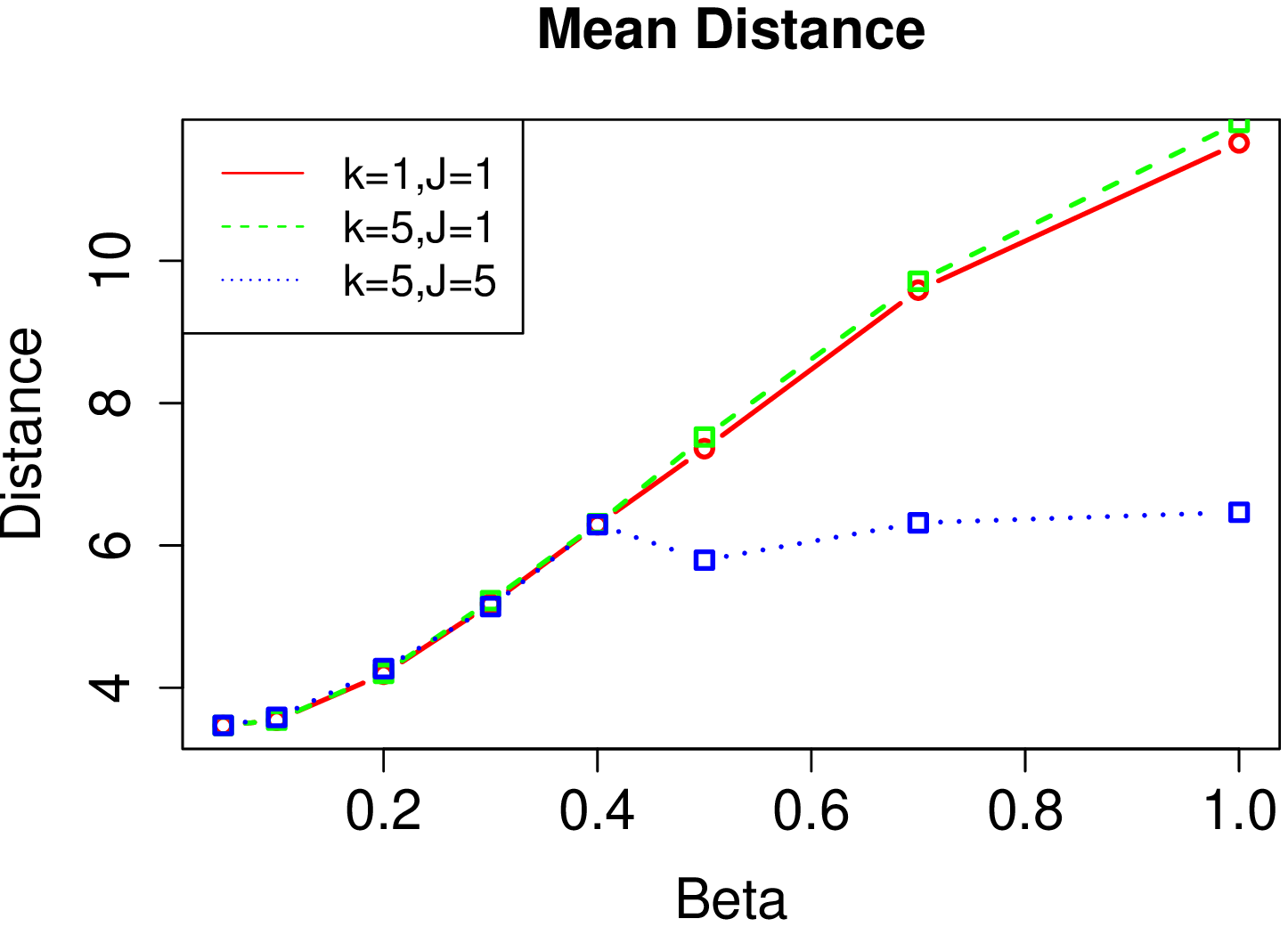,width=0.45\linewidth,clip=} &
		\epsfig{file=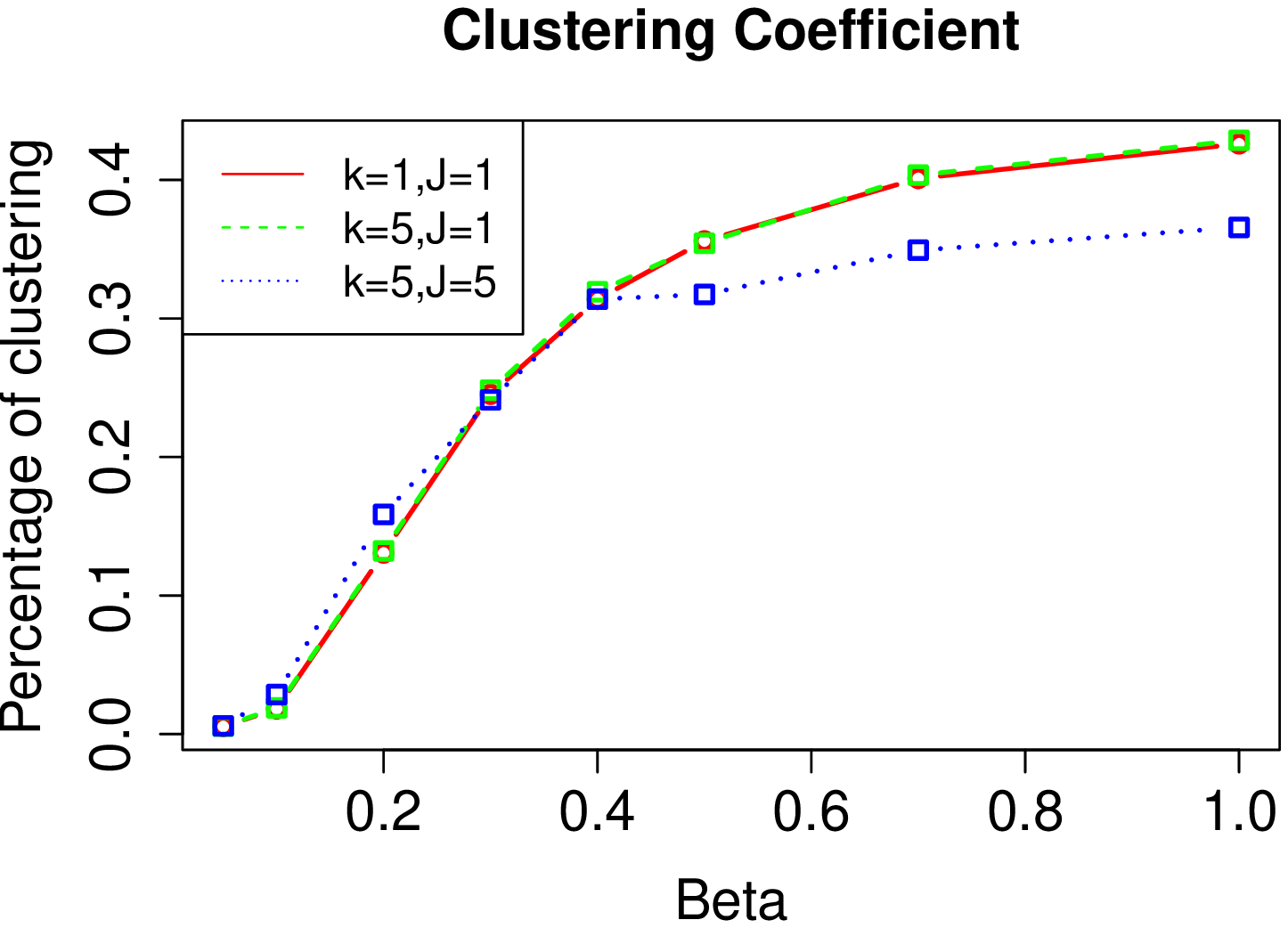,width=0.45\linewidth,clip=}
	\end{tabular}
	\caption{Network properties as a function of the temperature. All cases correspond to an average over 10 realizations starting with $d=2$ lattice with second neighbors connected and empty boundary conditions. {\it Left} Mean distance between the nodes. {\it Right} Clustering, measured as the proportion of adjacent nodes of each node that are connected.}\label{fig:n50kj}
\end{figure}

A series of simulations were run to test the model for different values of the parameters. Figure \ref{fig:n50kj} shows network properties as a function of $\beta$, mean distance between the nodes and the clustering coefficient (defined here as the percentage of adjacent nodes of each node that are connected) for some of those runs. All cases started from a regular $d=2$ lattice square with $n=50\times 50$ nodes with empty boundaries and second level neighbors. Each point in the graphics correspond to an average over 10 realizations. The observed variation of the results, measure by the standard deviation of each statistics was small enough to justify running only 10 cases of each set of parameters. Indeed, standard deviations for the clustering coefficients were usually at the $0.001-0.005$ range and, for the mean, in the $0.01-0.09$ range and were not included in the figures to avoid unnecessary cluttering.

We can easily see that, as we change $k$ and $J$, the network properties seem to be basically unaffected up to $\beta=0.4$. And, while for $\beta /geq 0.5$, we see a difference in behavior when $J=5.0$. However, comparing the cases where $J=1.0$ and only $k$ changes from 1 to 5, we see there is basically no observable difference. That is not surprising, as changing $k$ only means opinions get updated more often. Larger $k$ main expected effect is, as more opinion updates happen, the final values of $\nu$ become proportionally larger. That might have an influence on the network structure, but not necessarily. On the other hand, as $J$ is part of the energy $H$, it has a role on whether each attempted rewiring will succeed or not. It would be expected, therefore, that by changing $J$, we should observe an influence on the network statistics. That only happens, however, for larger values of $\beta$.

On the other hand, the tendency to connect agents who share the same preferences might not be so important when choosing between close nodes, in special for high temperatures. That lack of influence might be related to the fact that opinions tend to form local domains. As the spatial component works to make local connections more likely than distant ones, both effects might play similar roles. We do see that, for larger $J$, as the temperature decreases, differences start to appear. Instead of slowly becoming similar to the regular lattice that minimizes the distance component of the energy, the system seems to keep some amount of random characteristics due to stronger opinion  effects. Indeed, as $J=5$, the change in energy in Equation~\ref{eq:energydistop} one observes from disconnecting from a disagreeing neighbor and connecting to an agreeing one, $\Delta H = -2 \beta J = -10$ is large enough that even if the new neighbor distance is 10 units larger, the change in energy will be zero. The rewiring, in that case, will always be accepted. On the other hand, rewiring an agreeing edge to a disagreeing one becomes highly unlikely, except for large temperatures. And, of course, if there is no difference in the opinion part of the equation, only the distance will matter.

To check what might happen if opinions become much stronger, the case where we make $J=20$ was run, both for $\beta=0.05$ and $\beta=0.4$. In both cases,$k$ was kept at $k=1$. When $\beta$ is small, there is again no observable difference. That is expected as large temperatures correspond to an almost random network structure ($\beta =0$ would mean all changes are accepted). However, when $\beta=0.4$, both the observed mean distance ($4.29$) and the clustering coefficient ($0.15$) average results over 10 runs diverged importantly from the cases where $J=1$ or $J=5$. Comparing with the results shown in Figure~\ref{fig:n50kj} we can see that the decrease in temperature has a smaller effect on the system than it had for smaller $J$. By looking only at those figures, it seems the systems starts to organize but remains close to a random phase. Therefore, it makes sense to take a look at the configurations that are produced by the algorithm.

\begin{figure}
	\centering
	\begin{tabular}{cc}
		\epsfig{file=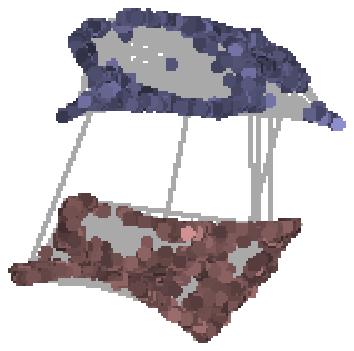,width=0.40\linewidth,clip=}&
		\epsfig{file=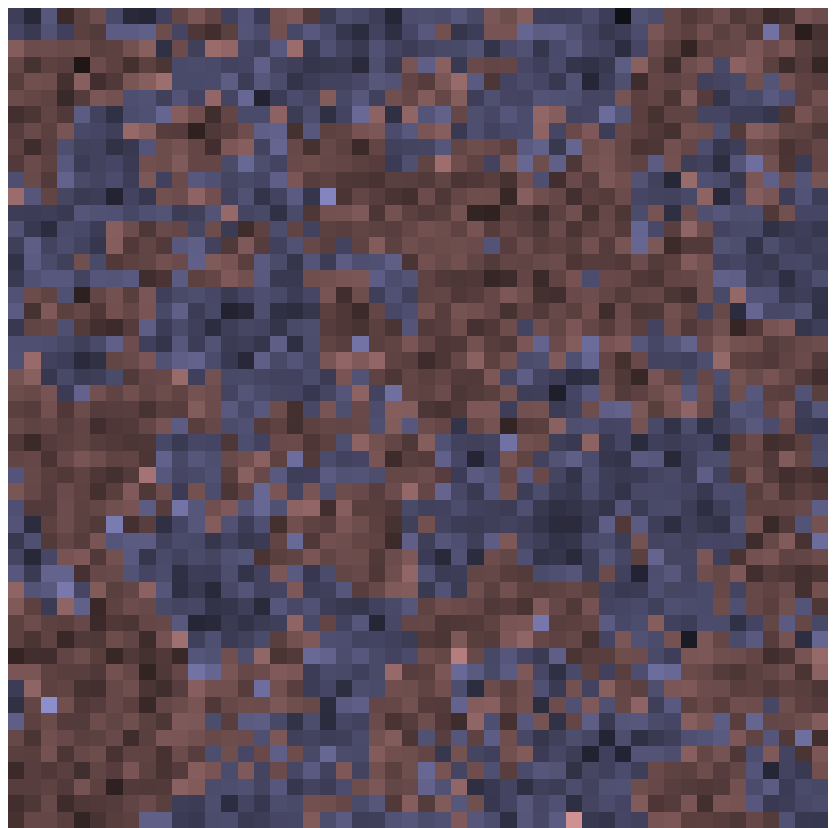,width=0.45\linewidth,clip=}
	\end{tabular}
	\caption{Typical network configuration generated after 100 average interactions per agent with $\beta=1.0$, starting with a $d=2$, $50\times 50$ lattice with second neighbors connected and empty boundary conditions. Blue and red are used to represent the two choices, lighter shades identifying less extreme opinions. {\it Left}Network structure drawn by the igraph package. {\it Right} Domain configuration with choices shown as a function of spatial position.}\label{fig:j20}
\end{figure}

Tests for a longer run still with $J=20$ were performed. Figure \ref{fig:j20} shows the final configuration for one run where edge are tentatively rewired an average of 100 times per node. The scenario in the figure corresponds to a low temperature ($\beta=1.0$) and $k=1$, with other parameters identical to the previous runs ($d=2$, $50\times 50$ lattice with second neighbors connected and empty boundary conditions). It is easy to observe in the left panel that the network is split between the two opinions, with a limited number of edges still connecting disagreeing agents. But the structure inside each agreeing region is also interesting. Distant agents find it hard to connect, even if they agree. As a consequence, unlike what the previous results suggested, each region is not close to a random graph. Instead, only closer agents tend to be connected. Even inside agreeing domains, we still observe local neighborhoods that connect to other agreeing regions less often than they would in a random situation. The choices on a lattice arranged according to the spatial $x$ and $y$ coordinates for the same run can be seen in the right panel of Figure \ref{fig:j20}. Despite the clear separation in the generated network, the opinions do not show the same spatial separation. Local clusters still appear but they do not grow big enough. That means several nodes with the same preference have disagreeing clusters between them. That fact contributes to make distances inside each agreeing domain larger. As a consequence, connections between distant clusters become more unlikely. Instead, inside each domain, the nodes still connect influenced by the spatial dependence. That explains the observed holes in the top panel.

\begin{figure}
	\centering
	\begin{tabular}{cc}
		\epsfig{file=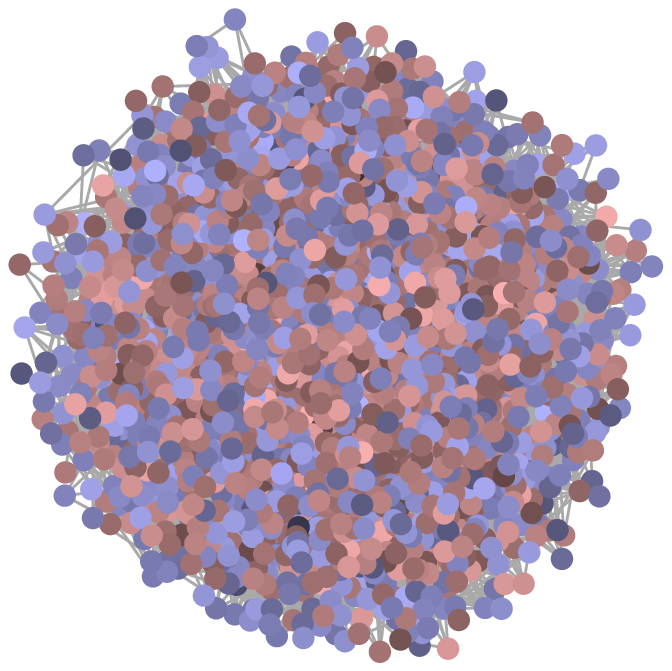,width=0.45\linewidth,clip=} & 
		\epsfig{file=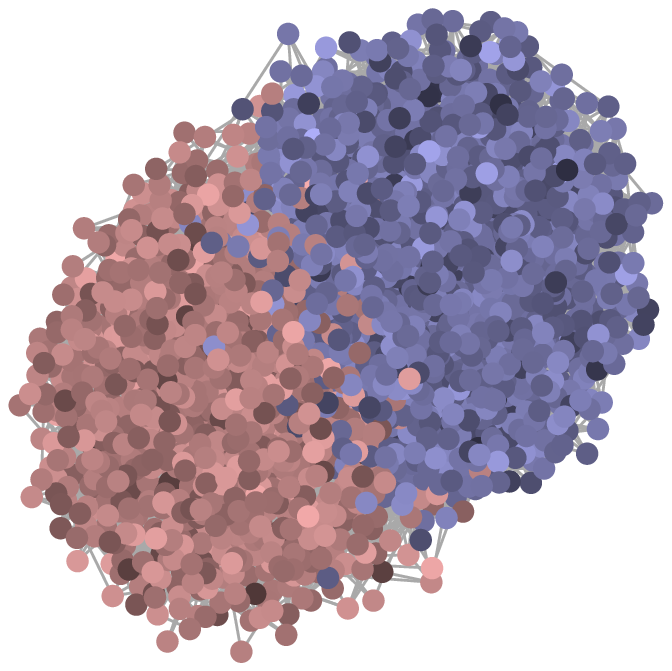,width=0.45\linewidth,clip=} \\
		\epsfig{file=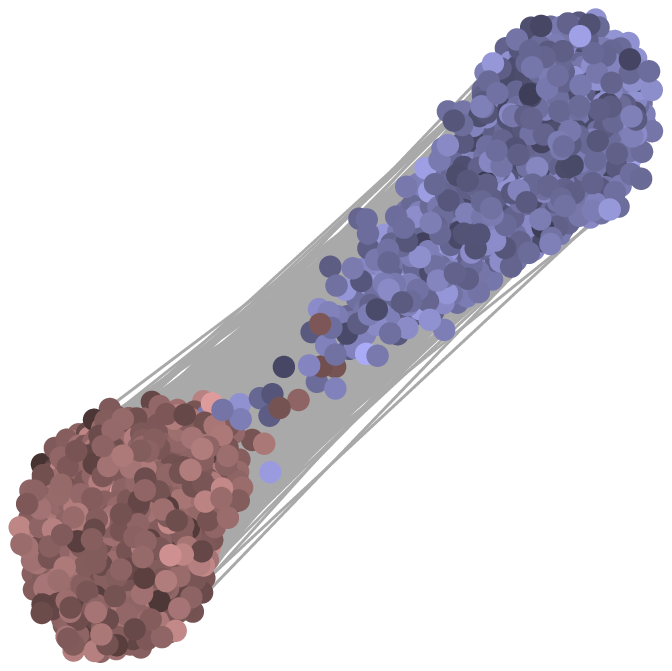,width=0.45\linewidth,clip=} &
		\epsfig{file=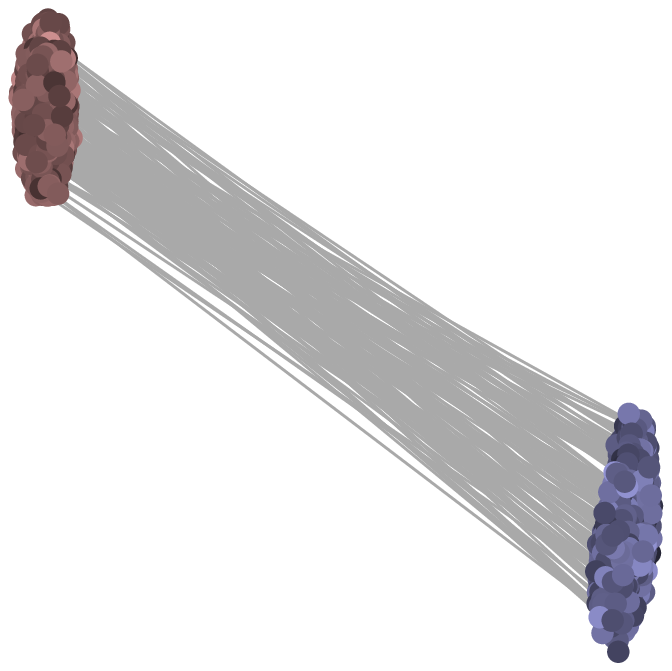,width=0.45\linewidth,clip=}
	\end{tabular}
	\caption{Typical final network configuration generated after 50 average interactions per agent for $k=1$ and $J=5$. $\beta$ values were chosen to represent different final configurations. {\it Top Left} $\beta = 0.1$. {\it Top Right} $\beta = 0.15$. {\it Bottom Left} $\beta = 0.2$. {\it Bottom Right} $\beta = 0.5$.  }\label{fig:sqrtdistj5}
\end{figure}

\begin{figure}
	\centering
	\begin{tabular}{cc}
		\epsfig{file=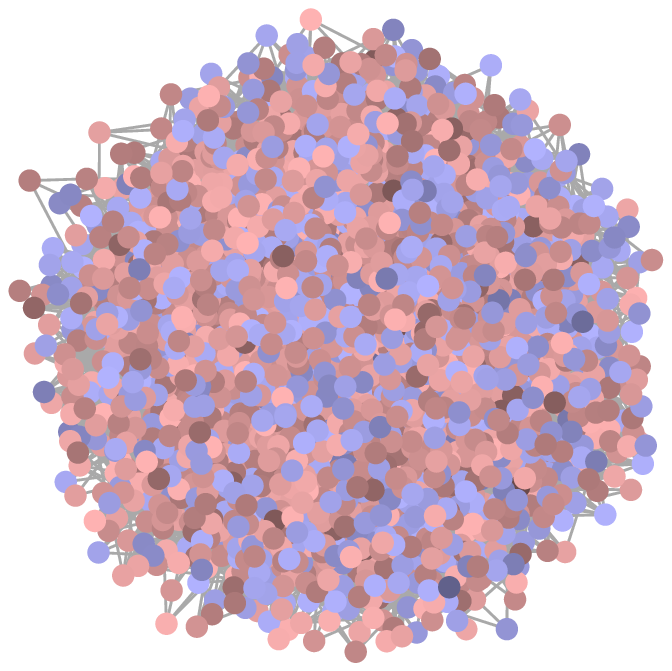,width=0.45\linewidth,clip=} & 
		\epsfig{file=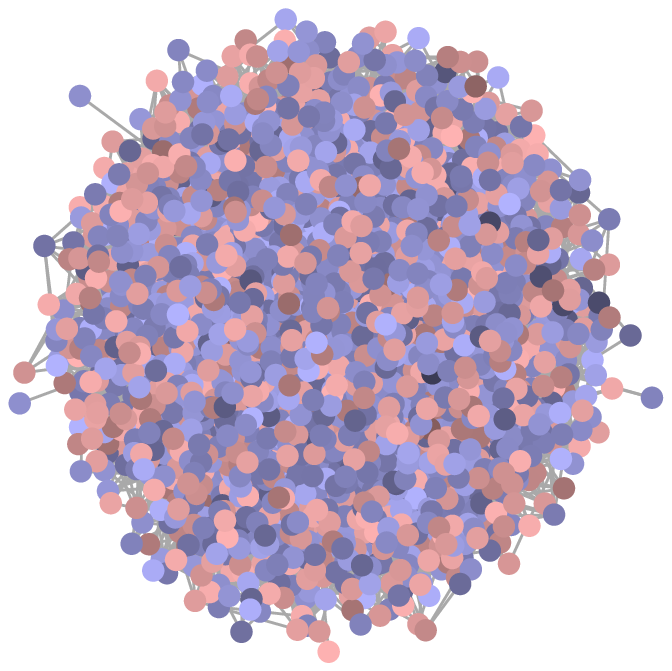,width=0.45\linewidth,clip=} \\
		\epsfig{file=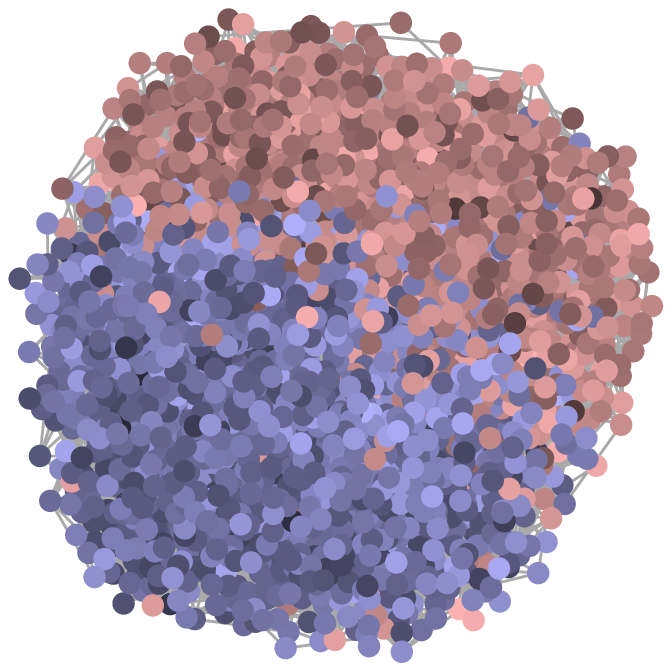,width=0.45\linewidth,clip=} &
		\epsfig{file=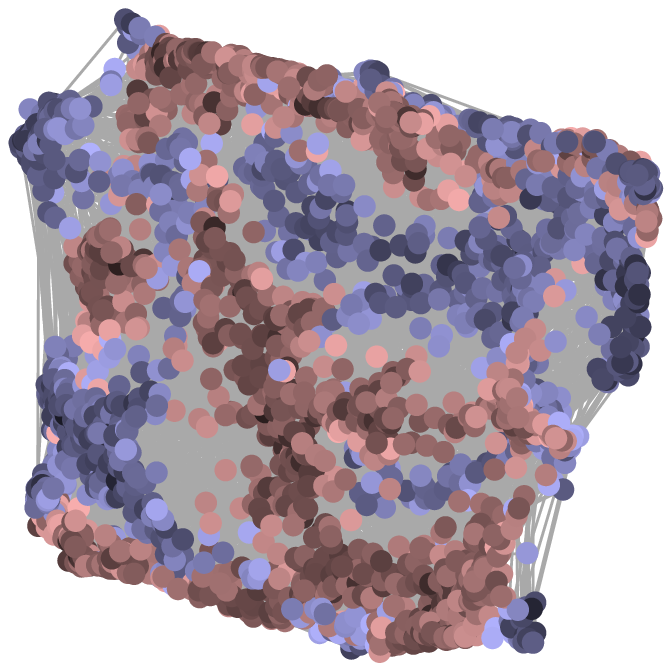,width=0.45\linewidth,clip=}
	\end{tabular}
	\caption{Typical final network configuration generated after 50 average interactions per agent for $k=1$ and $J=1$. $\beta$ values were chosen to represent different final configurations. {\it Top Left} $\beta = 0.1$. {\it Top Right} $\beta = 0.2$. {\it Bottom Left} $\beta = 0.5$. {\it Bottom Right} $\beta = 1.0$. }\label{fig:sqrtdistj1}
\end{figure}

\begin{figure}
	\centering
	\begin{tabular}{cc}
		\epsfig{file=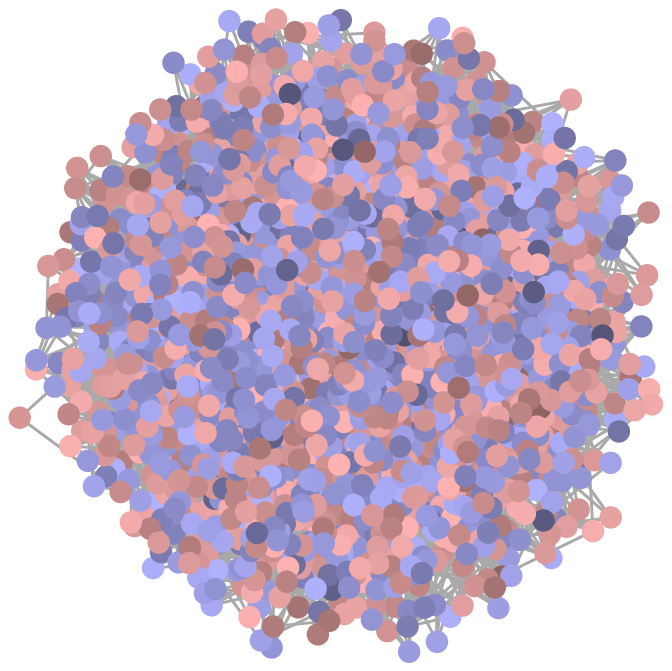,width=0.45\linewidth,clip=} & 
		\epsfig{file=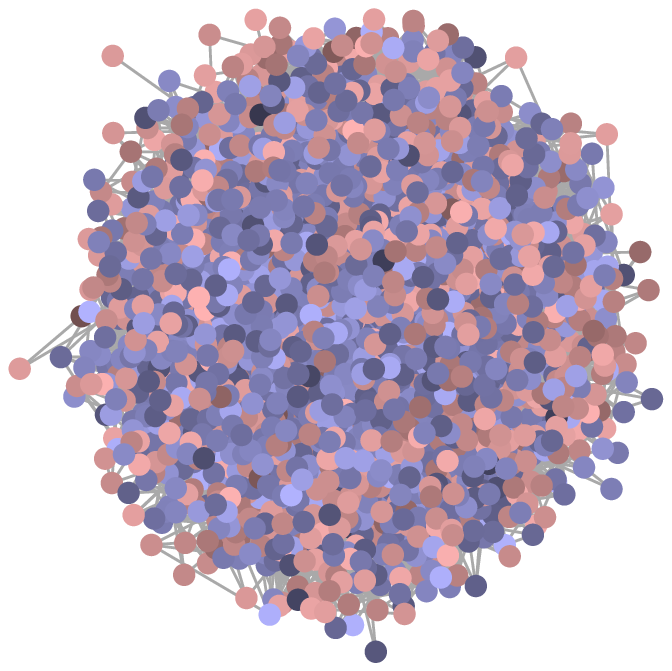,width=0.45\linewidth,clip=} \\
		\epsfig{file=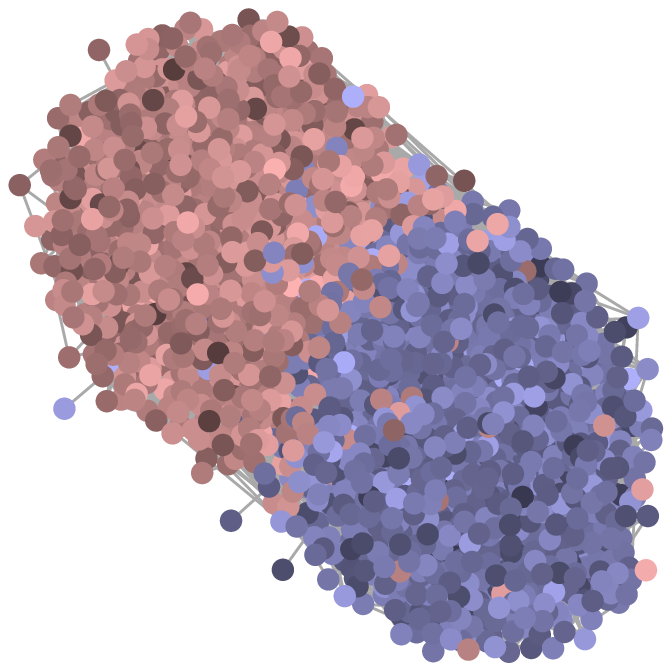,width=0.45\linewidth,clip=} &
		\epsfig{file=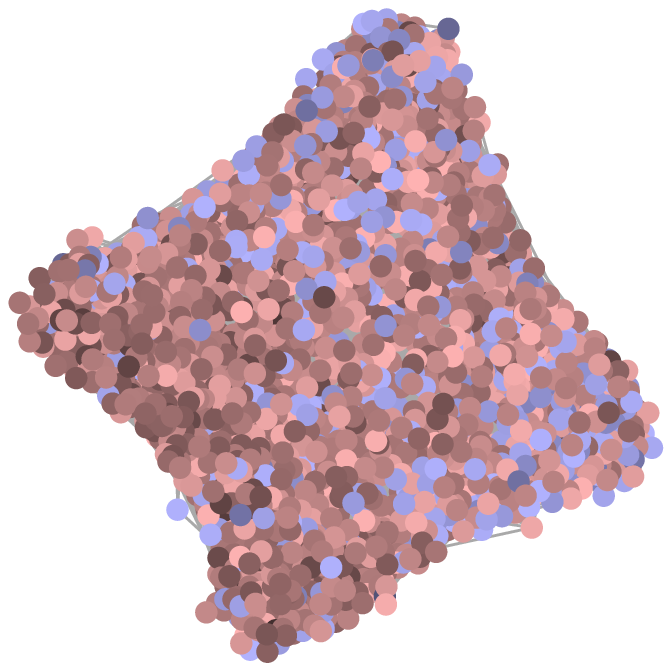,width=0.45\linewidth,clip=}
	\end{tabular}
	\caption{Typical final network configuration generated after 50 average interactions per agent for $k=1$ and $J=1$. $\beta$ values were chosen to represent different final configurations. Initial network state was random with the same number of vertices as previous cases. {\it Top Left} $\beta = 0.2$. {\it Top Right} $\beta = 0.5$. {\it Bottom Left} $\beta = 1.0$. {\it Bottom Right} $\beta = 2.0$. }\label{fig:sqrtdistj1BF}
\end{figure}

As another example of how $J$ becomes much more relevant when it is comparable to the maximum energy contribution to the energy in Equation~\ref{eq:energydistop}, a few simulations were also run with $\gamma=1/2$. That is, when distance contribution to the energy only grows with the square root of the actual physical distance. Final configurations for specific runs with $\gamma=1/2$ can be seen for different values of $\beta$ at Figures \ref{fig:sqrtdistj5}, \ref{fig:sqrtdistj1} and \ref{fig:sqrtdistj1BF}. Notice that the actual $\beta$ values do change between figures. Once more, all cases started  with a $d=2$, $50\times 50$ lattice with second neighbors connected and empty boundary conditions, except for Figure \ref{fig:sqrtdistj1BF} where initial network was random, with the same number of edges as in the other cases.

We can observe the transition from randomness of networks for small values of $\beta$ to the more ordered case of larger $\beta$s. Those transitions, however, do not happen equally for each case. As it should be expected, when $J$ is larger, as in Figure~\ref{fig:sqrtdistj5}, separation between the two domain preferences is observed for smaller values of $\beta$. Already at $\beta=0.15$, we can observe clear domains, even if they are linked by a large number of edges. At $\beta = 0.2$ those domains move further apart, but we still see some nodes (in that case, mostly blue ones) that are well connected to both domains and, therefore, appear in the transition zone. As $\beta=0.5$ the separation becomes very clear, with fewer edges between the two domains.

As we observe results for $J=1$ in Figure \ref{fig:sqrtdistj1}, there are important differences in how the system responds to a decrease in temperature. Both $\beta=0.1$ and $\beta=0.2$ seem visually to be close to random results. Since opinions have a much weaker influence on the update rule, that is to be expected. Even as $\beta=0.5$ the two domains start to separate but they are still mixed together. Interestingly, however, as $\beta$ increases even more, we do not observe the same splitting in two well defined domains we see when $J=5$. Instead of that, the spatial component of the energy shows its importance. Domains of blue and red are clearly observed, but they have a strong spatial localization as well. Indeed, the generated network almost seems to keep the initial regular lattice configuration it lost for smaller values of $\beta$.

That regular lattice resemblance could be an artifact of initial conditions, however. Each run does start with the network defined as a regular lattice. Therefore, it is worth asking if what we observe in Figure \ref{fig:sqrtdistj1} is not just an effect of those initial conditions and the fact we are not allowing the algorithm to run long enough to achieve thermal equilibrium. In order to check that possibility, Figure \ref{fig:sqrtdistj1BJ} shows the same choice of parameters as in Figure \ref{fig:sqrtdistj1}, except that the initial network is taken to be random with the same number of edges. The same basic behavior is observed, except that now larger values of $\beta$ are necessary for the networks to organize. As the system was not at its thermal equilibrium, it makes sense that larger values of $\beta$ would be required to compensate for the initial randomness. But we see that, even starting from randomness, the network eventually organizes itself according to the spatial position of the nodes as $\beta$ gets larger.

\begin{figure}
	\centering
	\begin{tabular}{cc}
		\epsfig{file=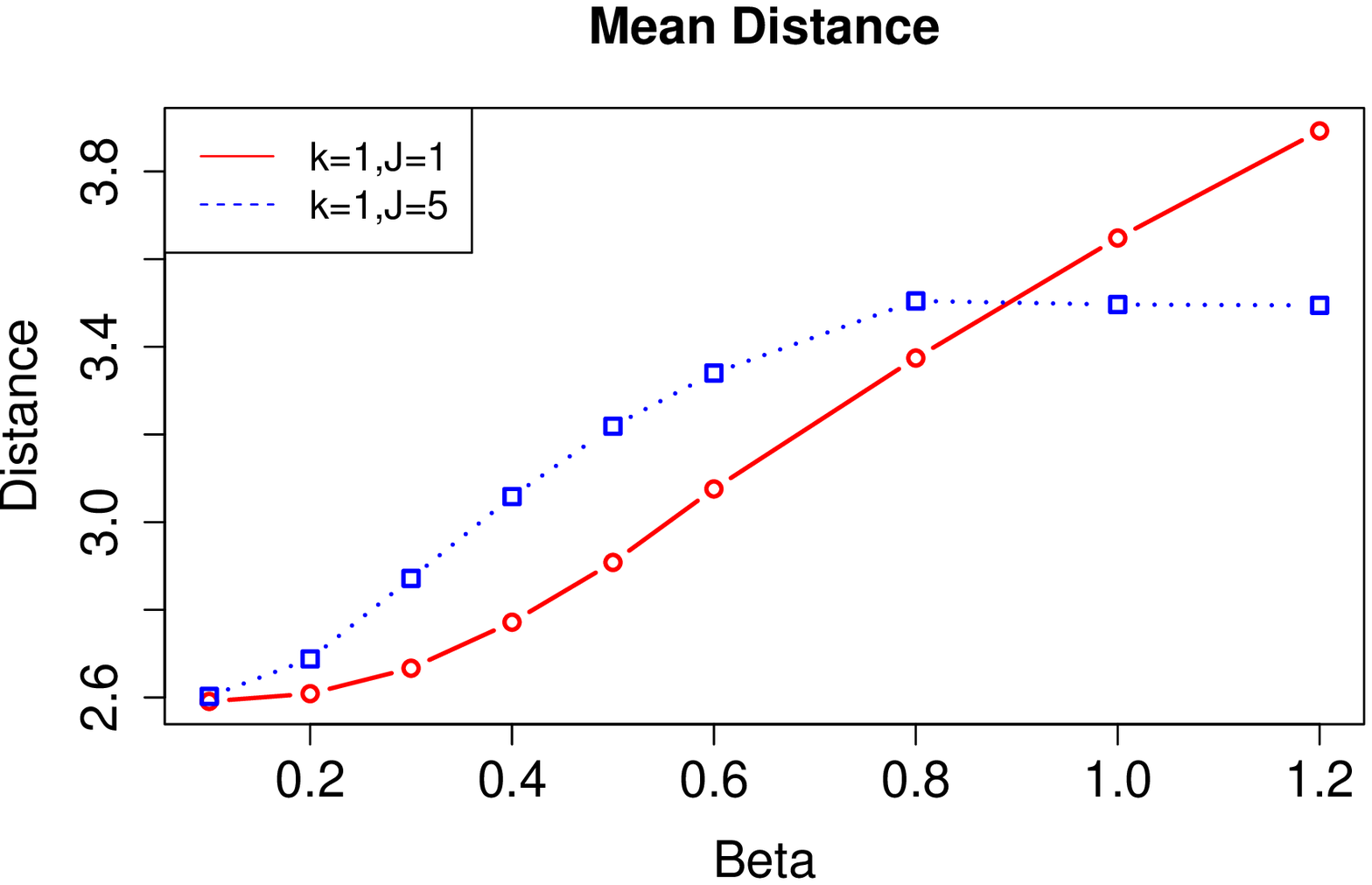,width=0.45\linewidth,clip=} &
		\epsfig{file=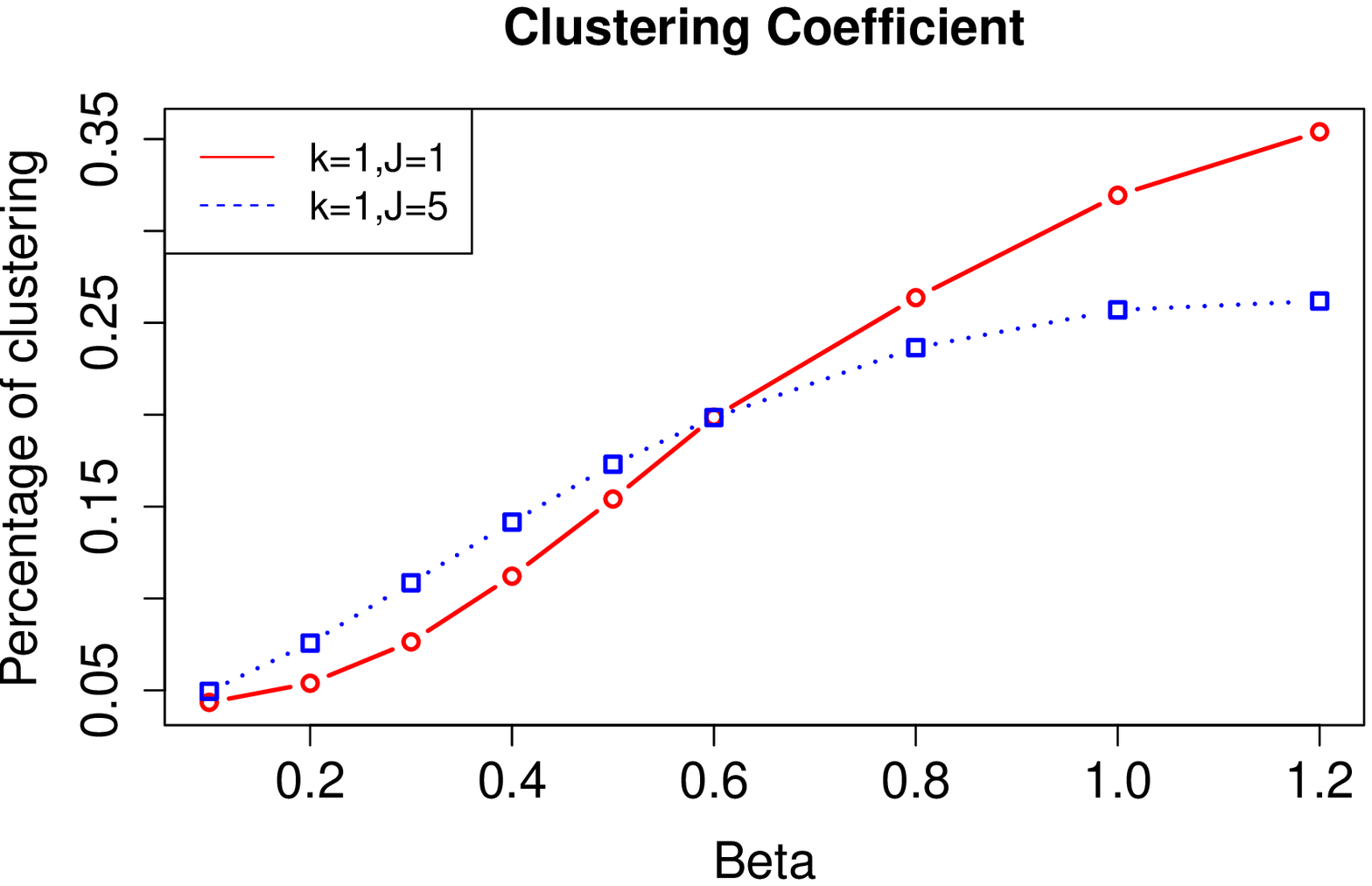,width=0.45\linewidth,clip=}\\
		\epsfig{file=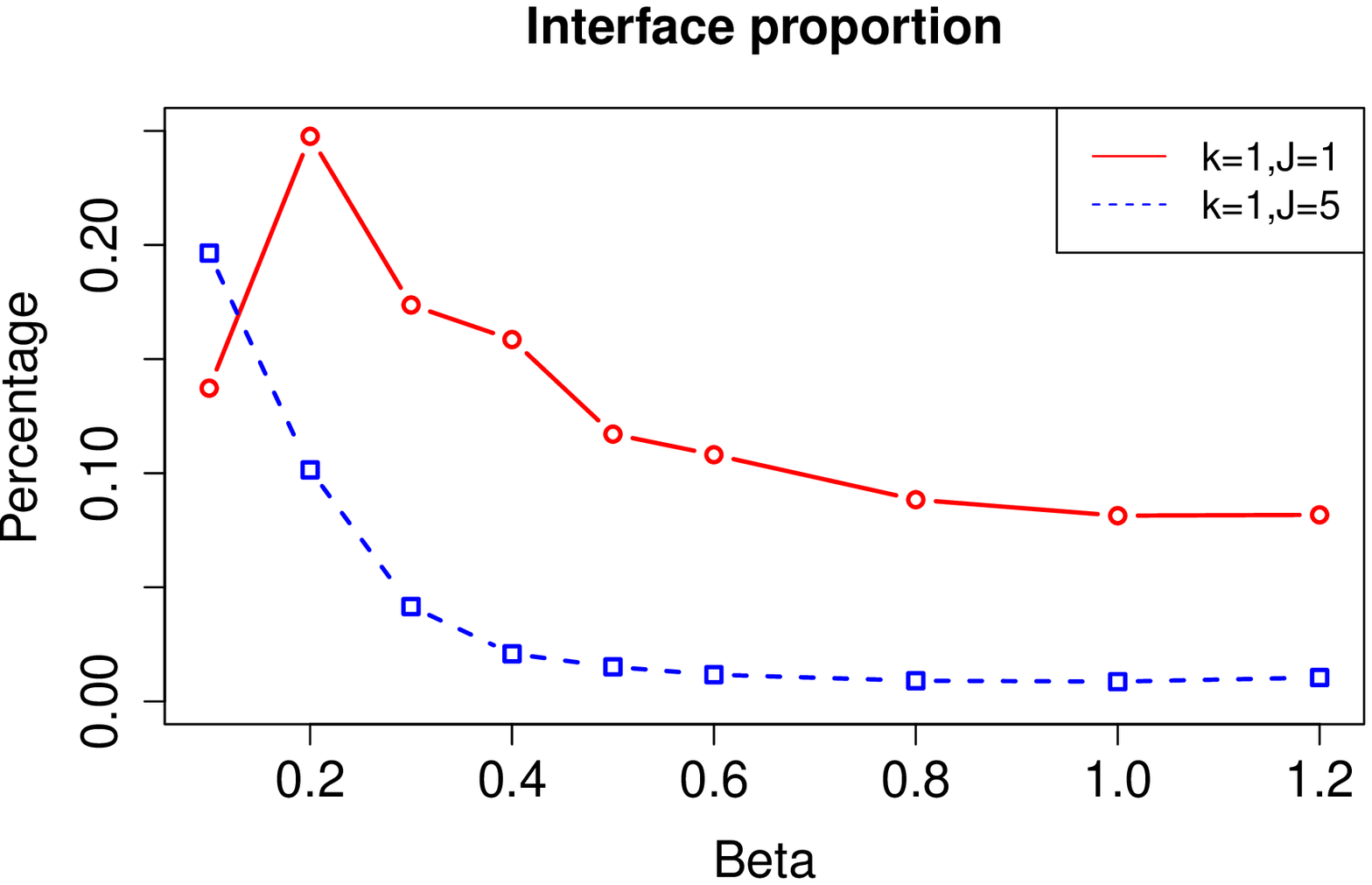,width=0.45\linewidth,clip=}
	\end{tabular}
	\caption{Network properties as a function of the temperature. All cases correspond to an average over 20 realizations starting with $d=2$ $16 \times 16$ lattice with second neighbors connected and empty boundary conditions. {\it Top Left} Mean distance between the nodes. {\it Top Right} Clustering, measured as the proportion of adjacent nodes of each node that are connected.{\it Bottom} Interface between the domains measured by the proportion of edges connecting nodes with opposing choices.}\label{fig:n16}
\end{figure}

In order to understand the behavior of the model at small scales, runs for a $d=2$ $16 \times 16$ lattice with second neighbors connected and empty boundary conditions were also performed. Each point in each graphic corresponds to the average over 20 implementations of the observed statistics at the end of the run. The network mean distance and the clustering coefficient can be seen in Figure~\ref{fig:n16} at the top left and top right panel respectively. To observe how well networks split in two domains for each case, the proportion of edges that correspond to disagreeing agents at each node was also calculated as a measure of the relative importance of the interface to the full network. Those results are also shown, at the bottom panel. Surprisingly, for $\beta=0.1$, the interface proportion was larger for $J=5$ than for $J=1$. In both cases, the standard deviation associated of the observed parameters was unusually high when compared with other situations, in the 4-5\% range. The standard deviation for the interface proportion did decrease with larger values of $\beta$s.

For $J=1$ both the mean distance and the clustering coefficient increase as the temperature drops and the system tends to the distance-optimal regular lattice. However, for $J=5$ that increase is halted for larger $\beta$s.  The system does move away from the random structure as temperature drops, as it should. However as the system reaches a state with clear opinion domains, the statistics show a tendency to stabilize and the regular lattice structure is no longer a consequence of low temperatures. As expected, $J$ has an important influence on the appearance of opinion domains. While the influence terms means that far less nodes correspond to a disagreement between nodes, for $J=1$, even at low temperatures, the system still keeps around 10\% of the nodes as interface ones. For $J=5$, however, the interface becomes very small, with proportions ranging in the 1-2\% interval.

\begin{figure}
	\centering
	\begin{tabular}{cc}
		\epsfig{file=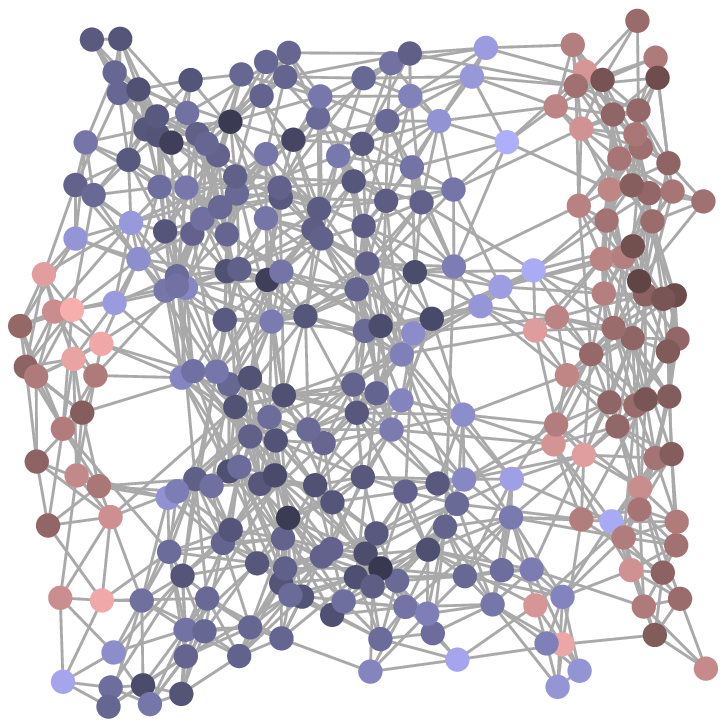,width=0.45\linewidth,clip=} & 
		\epsfig{file=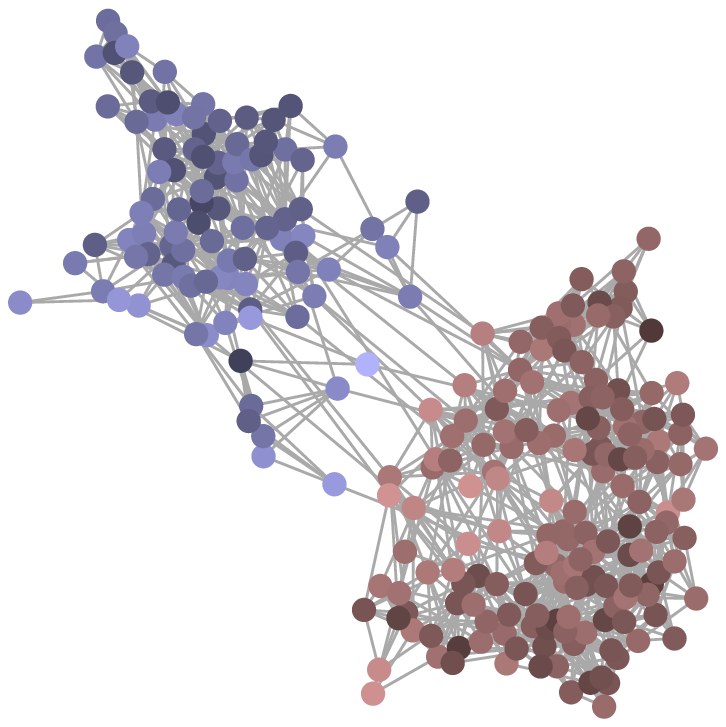,width=0.45\linewidth,clip=}
	\end{tabular}
	\caption{Typical final network configuration generated after 50 average interactions per agent from an initial $16 \times 16$ regular bi-dimensional lattice with no boundary conditions. In both cases, $\beta=1.0$ and $k=1$. {\it Left} $J=1$.  {\it Right}. $J=5$. }\label{fig:n16networks}
\end{figure}

Figure~\ref{fig:n16networks} shows clearly the difference between a $J=1$ and a $J=5$ network for a case where $\beta=1.0$. In the left panel we can see the regular $16 \times 16$ lattice structure imposed by the distance term in the energy. For a more important influence of opinions, as in the right panel, that structure is no longer visible. In both cases, most edges correspond to agreement as the opinion domains are clear. But, as the preference of the agents become more important, we see the fact there is a distance term in the energy might not be easy to identify from observing the network structure.

\section{Conclusions}

Implementing an algorithm where opinions have influence on the network of influences between agents can lead to a variety of connection types. Even when the agents start at an artificially ordered regular lattice, the distance structure might disappear. That can happen trivially due to high temperatures. And it can also happen because of the importance of opinions on the energy term. Echo-chambers, even when distance matters, can appear as a consequence of the dynamics. We also observe cases where clustering coefficients show that communities are kept, depending on the parameters. Even more general network descriptions might appear by introducing new parameters at the energy function. One obvious case that was not followed here is how the energy might depend on the degrees of the connecting nodes. In that scenario, studying the degree distributions of the generated network should be an interesting problem. If energy favors connections between nodes with larger degrees we might generate systems where some sort of preferential attachment happens.

The model in this paper can be used to create a fixed network based on opinions and network statistics. It can also provide a dynamical description of such a system. That can be fundamental in many situations. An analysis of the evolution of partisanship on 86 countries show that temporal effects can be important to understand that problem \cite{Michelitch2018}. Partisanship is an opinion problem. Therefore, a proper energy function for this situation would should include an opinion term. Also, adding a spatial component makes sense when we study systems where people convene for political reasons coming from different geographical regions where they have their base of operation and personal connections. Loyalty to one's party and negative affect towards the opposing one has been increasing in USA politics~\cite{Horwitz2015}. An analysis of the evolution of partisanship in the USA House of Representatives \cite{Abramowitz2016,Andris2015} has shown the clear appearance of party (in other terms, opinion) domains in the interaction between its members. More than that, the observed networks show the same splitting structure we obtained here. That suggests the approach proposed here can help better understand those systems.

\section{Acknowledgement}
The author would like to thank Funda\c{c}\~ao de Amparo \`a Pesquisa do Estado de S\~ao Paulo (FAPESP), for the support to this work, under grant 

\bibliographystyle{unsrt}
\bibliography{biblio}

\end{document}